\documentclass[reprint,amsmath,amssymb,aps,showpacs,prb]{revtex4-1}

\usepackage[dvips]{graphicx}
\usepackage{bm}
\usepackage{dcolumn}
\begin{document}


\title{Coherent reaction of Fermi superfluid on correlated disorder}


\author{I. A. Fomin}
\email{fomin@kapitza.ras.ru}
\affiliation{P. L. Kapitza Institute for Physical Problems Russian Academy of Science \\
Kosygina 2,
 119334 Moscow, Russia}

\author{E. V. Surovtsev}
\affiliation{P. L. Kapitza Institute for Physical Problems Russian Academy of Science \\
Kosygina 2,
 119334 Moscow, Russia}


\date{\today}

\begin{abstract}
Motivated by the disparity between the experimentally observed
properties of superfluid $^3$He in aerogel and predictions of the
Abrikosov and Gorkov theory of superconducting alloys we consider
effect of correlated pair-breaking impurities on the temperature
dependence of the square of magnitude of the average order parameter
and related thermodynamic properties of a Fermi superfluid. We show
that the correlations, increasing the spectral density of the
fluctuations with small wave vectors increase the transition
temperature and enhance a temperature region below $T_c$ where
fluctuations of the order parameter dominate. Outside of this region
we treat effect of impurities as a perturbation and express
corrections to thermodynamic properties in terms of the structure
factor of impurities.  Assuming a simple model expression for
correlation function of aerogel we find  corrections to the
temperature dependence of the NMR frequency shifts  for A-like and
B-like phases of $^3$He and compare these with experimental data.
\end{abstract}


\maketitle

\section{Introduction}
Effect of impurities on traditional superconductors is well
understood within the Abrikosov and Gor`kov (AG) theory of
superconducting alloys.\cite{AG} For some of more recent
superconducting and superfluid Fermi systems, including high-Tc
compounds and superfluid $^3$He, application of this theory is not
always justified. Superfluid $^3$He in high porosity silica
aerogel\cite{parp1,halp1} is a good example of such situation.
Thermodynamic properties of this object show both quantitative and
qualitative discrepancies with the generalization of the AG theory
for the p-wave Cooper pairing.\cite{thuneb} The observed difference
between the temperature of superfluid transition in the bulk liquid
and in aerogel $T_{cb}-T_{ca}$ at pressures above 20 bar is about a
half of the value, predicted by the theory.\cite{halp2} As an
example of qualitative discrepancy we can cite the anomaly in the
temperature dependence of the superfluid density and of the
magnitude of transverse NMR frequency shift. In Ginzburg and Landau
region both quantities instead of the linear growth with $T_{ca}-T$
demonstrate the dependence which can be approximated by a power law
$(T_{ca}-T)^n$ with $n\approx1.3-1.4$ and this dependence extends
well below the $T_c$.\cite{feff}

The model calculations of the structure of aerogel show that it is
an object, intermediate between the ensemble of conventional
impurities and a porous media.\cite{parp2,haard} On the one hand the
scattering centers having diameter $d\approx 3$~nm are much smaller
than the coherence length in $^3$He ($\xi_0\approx 20\div 80$ nm),
on the other -- the structure has low density regions (``voids'')
with the characteristic size of the order of $\xi_0$. For
improvement of the agreement with experiment more involved models
were suggested which combine elements of description in terms of
pores with that of impurities - isotropic inhomogeneous scattering
model (IISM),\cite{han} and its simplified version -
phenomenological IISM\cite{sauls2}, which is an interpolation
between the two limits  The models contain additional parameters
(only one in the simplified version). The use of these as fitting
parameters renders a satisfactory agreement between the calculated
properties of $^3$He and experimental data.\cite{halp2} Still these
models have certain shortcomings on a theoretical side. The
additional parameters are introduced on an intuitive basis without
precise definition of their relation to the structure of aerogel. As
a result  the values of these parameters can not be found
independently. It is not clear also how accurately predictions,
based on these models describe effect of real aerogel on  $^3$He.

In the present paper we also take into account inhomogeneity of
aerogel. Structural fluctuations, i.e. random deviations of the
local density of impurities from the average density are treated as
a perturbation. We concentrate on the effect of these fluctuations
on thermodynamic properties of the superfluid $^3$He near the $T_c$.
A starting approximation is the AG theory, where impurities are
treated as a continuous background, producing damping of
quasiparticles and characterized by one parameter -- mean free path
of quasiparticles~$l$. The first correction to the AG approximation
can be expressed in terms of the structure factor of aerogel, which
has precise definition and can be measured directly. The obtained
results are approximate, but their accuracy  is controlled by the
small parameter, which is a combination of parameters,
characterizing both aerogel and superfluid $^3$He.

For account of structural fluctuations of pair breaking impurities
we follow the argument of Larkin and Ovchinnikov.\cite{LO} They have
shown that on a low temperature side of the transition
$\langle\psi\rangle^2$ in addition to the term depending linearly on
$(T_c-T)/T_c$ acquires a singular correction proportional to
$\sqrt{(T_c-T)/T_c}$. At $T\rightarrow T_c$ this correction can
catch up with the principal term. That renders an estimation of a
region of ``broadening'' of the transition $\delta T_{br}$. At
$(T_c-T)\ll \delta T_{br}$ temperature dependence of $\rho_s$ and
other thermodynamic properties of a superfluid is determined by
fluctuations.\cite{feldm} For Gaussian fluctuations $\delta
T_{br}\sim(T_c/N_{\xi}^2)(\xi_0/l)^4$, where $N_{\xi}$ is a number
of scattering centers within a volume with the  radius $\sim \xi_0$.
In aerogel the cross-section of a scattering center is $\sim 10^2$
times greater than that of an atomic impurity in a metal. It means
that for the same mean free path number of centers in a unit volume
is $\sim 10^2$ smaller. Coherence length $\xi_0$ for $^3$He is of
the same order as in a typical superconductor, so the fluctuations
are bigger. Still for a 98\% aerogel $N_{\xi}\approx 50$ and
contribution of structural fluctuations dominates only in a region
$(T_c-T)/T_c\sim 10^{-4} - 10^{-5}$, which is too narrow to explain
the observed anomaly.

Porto and Parpia\cite{parp2} pointed to the fact that in aerogel the
distribution of scattering centers is far from being random. They
gave convincing experimental evidences of strong effect of
structural correlations within the aerogel on the shift of the $T_c$
and on the temperature dependence of $\rho_s$. Guided by these
observations we consider theoretically effect of correlations in
distribution of pair-breaking  impurities on the shift of the T$_c$
and  on the  temperature dependence of the square of the absolute
value of the order parameter of a Fermi superfluid in a vicinity of
$T_c$. We introduce in the argument of Ref.~\onlinecite{LO} effect
of correlations. Because of the coherence of condensate its
interaction with the ensemble of impurities depends on phase shifts
between the perturbations brought about by the centers. The result
of interference brings in the numerator of the expression for
broadening of the transition an extra factor $\sim (nR^3)^2$, where
$R$ is the correlation radius. At sufficiently large $R$ this factor
compensates for  $N_{\xi}^2$ in the denominator and effect of
structural fluctuations can extend over an appreciable region below
the $T_c$ altering in this region temperature dependence of
thermodynamic properties of the superfluid.

 The paper is organized as follows: in Section \ref{scalar} we consider correlated impurities in a superconductor with a scalar
 order parameter. Short account of these results was published before.\cite{fom1,fom2} Here we added as illustration
 an example of a system with a realistic correlation function and dropped all results of Ref. \onlinecite{fom1}, which go
 beyond the perturbation theory and contain uncontrolled approximations. Reformulation of the argument for the p-wave
 superfluid $^3$He is presented in Section \ref{p-wave}.
 In Section \ref{comparison} comparison with the NMR data for ABM and BW phases of $^3$He is made.

\section{\label{scalar}Scalar order parameter}
In a vicinity of $T_c$ effect of a quenched disorder can formally be
described as a random spatial variation of coefficients in the
Ginzburg and Landau equation. For a scalar order parameter $\Psi$:
\begin{equation}
a(\textbf{r})\Psi+b(\textbf{r})\Psi|\Psi|^2-\nabla(c(\textbf{r})\nabla\Psi)=0.
\end{equation}
Following the procedure of Ref. \onlinecite{LO} we take into account
only spatial dependence of  $a(\textbf{r})$, because it renders the
most singular contribution to the thermodynamic properties at
$T\rightarrow T_c$.  For $b(\textbf{r})$ and $c(\textbf{r})$ we use
their average values $\langle b\rangle$ and $\langle c\rangle$. The
coefficient $a(\textbf{r})$ can be rewritten in terms of the local
transition temperature $T_c(\textbf{r})$:
$a(\textbf{r})=\alpha(T-T_c(\textbf{r}))$. It is convenient to
divide $T_c(\textbf{r})$ into its average value $\langle T_c\rangle$
and a relative fluctuation $\eta(\textbf{r})$: $a(\textbf{r})=
\alpha\langle T_c\rangle[\tau-\eta(\textbf{r})]$,
 where $\tau=-1+T/\langle T_c\rangle$ and
 $\eta(\textbf{r})=-1+T_c(\textbf{r})/\langle T_c\rangle$.
 After substitution $\Psi=\Psi_0\psi$ with
 $\Psi_0^2=\alpha\langle T_c\rangle/\langle b\rangle$, and
 $\xi_s^2=\langle c\rangle/\alpha\langle T_c\rangle$  Eq.~(1) takes the form
 \begin{equation}
 [\tau-\eta(\textbf{r})]\psi+\psi|\psi|^2-\xi_s^2\Delta\psi=0.
 \end{equation}
The random function $\eta(\textbf{r})$ is treated as a perturbation.
The global $T_c$ is defined by the condition $\langle\psi\rangle\neq
0$ at $T<T_c$. Below $T_c$ solution of Eq.~(2) can be sought in a
form $\psi(\bm{r})=\langle\psi\rangle(1+\chi({\bm{r}}))$ and the
perturbation procedure is justified when $\langle\chi^2({\bm
r})\rangle\ll 1$.  For real $\eta(\bm{r})$ essential part of $\chi$
is also real. Keeping in Eq.~(2) terms up to the second order in
$\chi$ and  $\eta$ we arrive at:
\begin{equation}
\tau-\eta+\tau\chi-\eta\chi+\langle\psi\rangle^2(1+3\chi+3\chi^2)-\xi_s^2\Delta\chi=0.
\end{equation}
By the definition $\langle\eta({\bf r})\rangle$=0 and
$\langle\chi({\bf r})\rangle$=0. Taking average of Eq.~(3) we arrive
at the equation for $\langle\psi\rangle^2$:
 \begin{equation}
\langle\psi\rangle^2=\frac{\langle\eta\chi\rangle-\tau}{1+3\langle\chi^2\rangle}.
\label{Ave_Eq}
\end{equation}
Subtraction of the averaged part from Eq.~(3) renders linear
equation for $\chi({\bf r})$:
\begin{equation}
(\tau-\eta({\bf r})+3\langle\psi\rangle^2)\chi-\xi_s^2\Delta\chi=
\eta({\bf r})-\langle\eta\chi\rangle.
\label{Fluc_Eq}
\end{equation} The second order term
$3\langle\psi\rangle^2(\chi^2-\langle\chi^2\rangle)$ is neglected in
comparison with $3\langle\psi\rangle^2\chi$.
Equation~(\ref{Fluc_Eq}) can be formally rewritten in terms of the
Green function $G({\bf r},{\bf r'})$:
\begin{equation}
\chi({\bf r})=\int G({\bf r},{\bf r'})(\eta({\bf r'})-
\langle\eta\chi\rangle)d^3r'.
\end{equation}
A straightforward argument shows that the average
$\langle\eta\chi\rangle$  can be expressed via the self-energy of
the averaged Green function of Eq.~(\ref{Fluc_Eq}) in the momentum
representation $\Sigma(\textbf{k},\tau)$:
\begin{equation}
\langle\chi\eta\rangle=\Sigma(0,\tau).
\end{equation}
In a principal order on the perturbation $\eta(\textbf{r})$
\begin{equation}
\Sigma(0,\tau)=\int\frac{\langle\eta(\textbf{-k})\eta(\textbf{k})\rangle}
{\tau+3\langle\psi\rangle^2-\Sigma(\textbf{k},\tau)+\xi_s^2k^2}
\frac{d^3k}{(2\pi)^3}.
\end{equation}
Now we take $u=\Sigma(0,\tau)-\tau$ as a new variable.
The self energy in the denominator of Green function may be expanded as $\Sigma(\textbf{k},\tau)\approx\Sigma(0,\tau)+\frac{\partial\Sigma}{\partial(k^2)}k^2$
and $\frac{\partial\Sigma}{\partial(k^2)}$ may be included in the definition
of $\xi_s^2$ Transition temperature is determined by the condition $u=0$.
In the second order on $\eta(\textbf{r})$ it is
\begin{equation}
T_{c2}=\langle T_c\rangle
[1+\int\frac{\langle\eta(\textbf{-k})\eta(\textbf{k})\rangle}
{\xi_s^2k^2}\frac{d^3k}{(2\pi)^3}]. \label{trans_temp_shift}
\end{equation}
Using these definitions and Eq.~(8) we arrive at the relation
between $u$ and $t=(T-T_{c2})/\langle T_c\rangle$:
\begin{equation}
u\left[1+2\int\frac{\langle\eta(\textbf{-k})\eta(\textbf{k})\rangle}
{\xi_s^2k^2[2u+\xi_s^2k^2]}\frac{d^3k}{(2\pi)^3}\right]=-t.
\label{u-equation}
\end{equation}
In evaluation of the average
$\langle\eta(\textbf{-k})\eta(\textbf{k})\rangle$ we neglect effect
of multiple scattering of quasiparticles  by the scattering centers.
That introduces a relative error in evaluation of $\langle
T_c\rangle$ of the order of $x^{2/3}$, where $x$ is the volume
concentration of impurities. Assuming also that all scattering
centers are identical we have:
$\eta(\textbf{r})=\sum_a\eta^{(1)}(\textbf{r}-\textbf{r}_a)-\eta^{(1)}(0)$
, where $\eta^{(1)}(\textbf{r}-\textbf{r}_a)$ is a local suppression
of transition temperature by one center situated at the position
$\textbf{r}_a$, $\eta^{(1)}(\textbf{k})$ -- its Fourier transform ,
$\eta^{(1)}(0)\equiv\eta^{(1)}(\textbf{k}=0)$. Then
$\langle\eta(\textbf{-k})\eta(\textbf{k})\rangle=
n|\eta^{(1)}(\textbf{k})|^2S(\textbf{k})$, where  the structure
factor $S(\textbf{k})=\langle\sum_b
\exp[i\textbf{k}(\textbf{r}_b-\textbf{r}_a)]\rangle$ depends on the
distribution of impurities. If impurities are not correlated all
terms with $\textbf{r}_b\neq\textbf{r}_a$ for finite $\textbf{k}$
vanish at the averaging. Then $S({\bf k})=1+(2\pi)^3n\delta({\bf
k})$.  Here unity comes from the summand with ${\bf r}_b={\bf r}_a$.
The term proportional to $\delta({\bf k})$ can be dropped because it
eventually enters with the factor
$\langle\eta(\textbf{r})\rangle=0$.   When correlations are present
terms with $\textbf{r}_b\neq\textbf{r}_a$ render finite contribution
to the structure factor. The number of additional terms in the sum
is of the order of $nR^3$, where $R$ is correlation radius. If
$nR^3\gg 1$ their sum, which in what follows is denoted as
$\widetilde{S}({\bf k})$ can be a principal contribution to $S({\bf
k})$. The $\widetilde{S}({\bf k})$ is  directly related to the
correlation function in the coordinate representation. For globally
isotropic distributions the probability $w({\bf r}_b|{\bf r}_a)$ to
find a particle in the point ${\bf r}_b$ if there is a particle in
the point ${\bf r}_a$ depends only on a distance $r=|{\bf r}_b-{\bf
r}_a|$. At $r\rightarrow \infty$ correlations vanish and $w(r)$
tends to a constant. Normalization of $w(r)$ is usually chosen so
that this constant is unity. Then a measure of correlations is
$v(r)=w(r)-1$.  Changing in the definition of $S(\textbf{k})$
summation for integration we arrive at
\begin{equation}
\widetilde{S}({\bf k})=n\int v(r)e^{-i{\bf k}{\bf r}}d^3r.
\label{Struc_Fac}
\end{equation}
Eq.~(\ref{u-equation}) can now be rewritten as
\begin{equation}
u[1+W(u)]=-t, \label{u_t_relation}
\end{equation} where
\begin{equation}
W(u)=2\int\frac{n|\eta^{(1)}(\textbf{k})|^2[1+\widetilde{S}(\textbf{k})]}
{\xi_s^2k^2[2u+\xi_s^2k^2]}\frac{d^3k}{(2\pi)^3}
\end{equation}
Numerator of the expression under the integral is the spectral
density of fluctuations. It is multiplied by the response function,
which is singular at  $u\rightarrow 0$ and $k\rightarrow 0$.
Contribution of fluctuations with $k\sim 1/\xi(u)$, where
$\xi(u)=\xi_s/\sqrt{2u}\gg\xi_s$ is strongly enhanced. The
perturbation  $\eta^{(1)}(\textbf{r})$ falls out on a distance $\sim
\xi_s$ from the center so that in the essential region of $k$
$\eta^{(1)}(\textbf{k})\approx\eta^{(1)}(0)\sim \xi_s\sigma$, where
$\sigma$ is a cross-section of scattering of a quasi-particle by one
center.  When  fluctuations are not correlated $\widetilde{S}({\bf
k})=0$, the spectral density is constant and we recover
 the result of Larkin and Ovchinnikov \cite{LO} for broadening of the transition:
 $W_1(u)=n|\eta^{(1)}(0)|^2/(2\pi\xi_s^3\sqrt{2u})$. Rough estimation renders
 $W_1(u)\sim (\frac{\xi_s}{l})^2\frac{1}{n\xi_s^3\sqrt{u}}$.
 For $^3$He in the 98\% aerogel $W_1(u)\ll 1$ if $u\geq 10^{-4}$.
 Contribution of correlations
\begin{equation}
W_2(u)=2n|\eta^{(1)}(0)|^2\int\frac{\widetilde{S}(\textbf{k})}
{\xi_s^2k^2[2u+\xi_s^2k^2]}\frac{d^3k}{(2\pi)^3}.
\end{equation}
depends on the behavior of  $\widetilde{S}(\textbf{k})$ at $k\sim
1/\xi(u)$.
 X-rays scattering data \cite{frel,parp2,haard} show that silica aerogels
 have fractal structure with the fractal dimension $D_f\approx1.7\div 1.9$
 depending on a sample. Formally it means that at $k\rightarrow 0$ the
 $\widetilde{S}(\textbf{k})$ grows as $1/k^D_f$ increasing
 density of the long wavelength fluctuations until it saturates
 at  $k\sim 1/R$.
 Further enhancement comes from the response function. It is characterized by the
 temperature dependent coherence length $\xi(u)$.  If
  $R\gg \xi_s$ there are two regions of $u$, corresponding to different asymptotic
  behavior of $W_2(u)$:  $\xi(u)>R$ and $\xi(u)<R$ with different asymptotic.
For finding the asymptotic  we express $W_2(u)$ in terms of the correlation
function $v(r)$:
\begin{equation}
W_2(u)=\frac{n^2|\eta^{(1)}(0)|^2}{\xi_s^2u}\int[1-\exp(-r/\xi(u))]
v(r)rdr. \label{sec-ord-cor}
\end{equation}

In a simplest case $v(r)$ can be characterized by two parameters --
a correlation radius $R$ and an overall amplitude $A$. Convenient
model form was suggested in Ref. \onlinecite{frel}:
\begin{equation}
v(r)=A\left(\frac{R}{r}\right)^{3-D_f}\exp(-r/R).
\end{equation}
This expression reproduces qualitative features of correlations in
silica aerogels i.e. the existence of a macroscopic correlation
radius $R$ and a fractal behavior with dimensionality $D_f$ at
$r<R$. Introduction of the dimensionless variable $y=r/R$ renders:
\begin{equation}
W_2(u)=\frac{n^2R^2|\eta^{(1)}(0)|^2}{\xi_s^2u}\int[1-\exp(-yR/\xi(u))]
v(y)ydy.
\end{equation}
Asymptotic of this integral at $(R/\zeta(u))\rightarrow 0$  is
$W_2(u)\approx
(n^2|\eta^{(1)}(0)|^2R^3\sqrt{2}/\sqrt{u}\xi^3)\int_0^{\infty}v(y)y^2dy$.
It has the same $1/\sqrt{u}$ singularity at $u\rightarrow 0$ as the
contribution of non-correlated impurities. Its sign and the relative
weight depend on the value of the integral
$I_v=\int_0^{\infty}v(y)y^2dy\sim A$. It means that the contribution
of correlations has an extra coefficient $AnR^3$.   For  realistic
values of parameters effect of fluctuations in aerogel is enhanced.

In the opposite limit $R\gg \xi(u)$ the exponent in the square
bracket under the integral in Eq.~(\ref{sec-ord-cor}) can be
neglected and
$W(u)\approx\frac{nR^2}{2\xi^2u}\int_0^{\infty}v(y)ydy$. That
renders a constant shift in the relation between $u$ and $t$ which
is equal to $(T_{c2}-\langle T_c\rangle)/\langle T_c\rangle$ so that
eventually $u\approx -\tau$. This is a straight line, which
extrapolates to $T=\langle T_c\rangle$ at $u=0$.

Relation (\ref{u_t_relation}) between $u$ and $t$ was derived with
the use of perturbation theory, it breaks down when $W(u)\sim 1$.
This condition determines a region of broadening of the transition
by structural disorder. Using the asymptotic of $W(u)$ at
$u\rightarrow 0$ we arrive at the estimation of the width of this
region: $u\leq (AR^3/\xi_sl^2)^2$.  The estimation coincides with
one obtained from the condition $\langle\chi^2\rangle\ll 1$. It
follows from the asymptotic at $R\ll\xi(u)$ of the  expression
\begin{equation}
\langle\chi^2\rangle=\int\frac{n|\eta^{(1)}(\textbf{k})|^2[1+\widetilde{S}(\textbf{k})]}
{[2u+\xi_s^2k^2]^2}\frac{d^3k}{(2\pi)^3}.
\end{equation}
The region of broadening of the transition for correlated impurities
is $(AnR^3)^2$ times more wide than for non-correlated and the
``tail'' of the singular correction to the linear dependence of $u$
on $t$ extends for a larger interval of $u$  into the region
$(AR^3/\xi_sl^2)^2\ll u \ll 1$ where perturbation theory holds.  If
$R\sim\xi_s$ the deviations of $\langle\psi\rangle^2$ from the
linear law can persist in a region well below the transition
temperature in agreement with the data for superfluid $^3$He in
aerogel.\cite{parp2,feff}

\begin{figure}[t]
\includegraphics[scale=0.37]{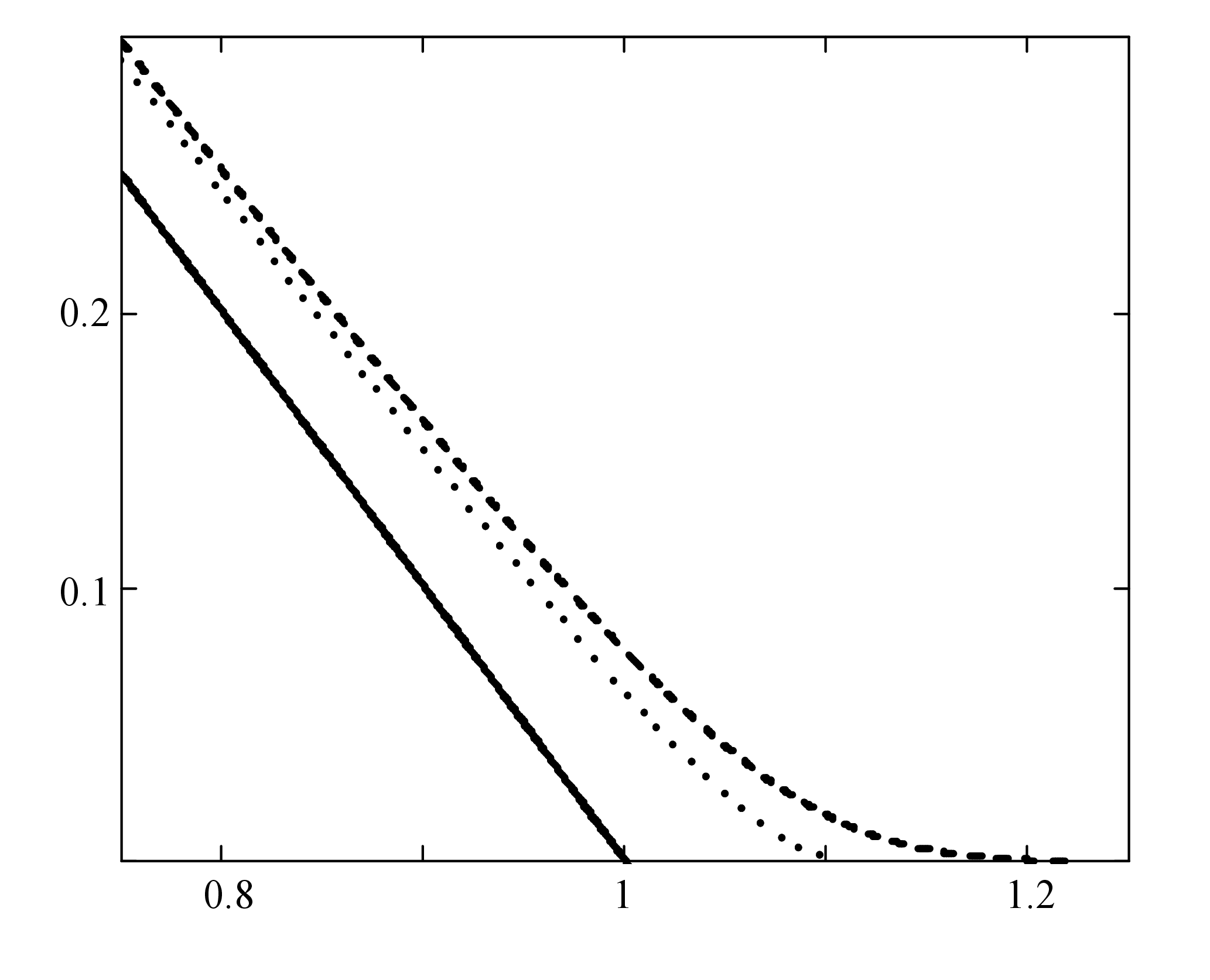}
\scalebox{0.9}[0.9]{\rotatebox{90}{$~~~~~~~~~~~~~~~~~~~~~~~~u(T/\langle
T_c \rangle)$}}
\scalebox{0.9}[0.9]{\rotatebox{0}{$~~~~~~~~~~~~~~T/\langle T_c
\rangle~~~~~~~~~~~$}}
\caption{\label{fig:epsart1} Dependence $u$ on $T/\langle T_c
\rangle$, as given by Eqns.~(12)~-~(16) for different ratios of
$R/\xi_s$ at a fixed value of the product $A(R/\rho)^{3-D_f}$. Solid
line corresponds to non-correlated case, i.e. $R/\xi_s=0$. Dotted
line corresponds to $R/\xi_s=3$ and dashed line is for $R/\xi_s=7$.}
\end{figure}

For illustration of qualitative changes, introduced by correlations
we plot on the Fig.~\ref{fig:epsart1} the dependence $u(T/\langle
T_c\rangle)$, given by Eqns.~(12)~-~(16) for different values of
parameters, characterizing aerogel. As one can see the transition
temperature shifts up with respect to  $\langle T_c\rangle$ when $R$
increases. Relative value of the shift according to
Eq.~(\ref{trans_temp_shift}) is of the order of $(R/l)^2$. Ratio of
the positive shift to $T_{cb}-\langle T_c\rangle$ is of the order of
$R^2/\xi_sl$. If $R\gg\xi_s$ a major change of the slope of $u(t)$
takes place at $u\sim(\xi_s/R)^2$.

Function $u(t)$ determines also correction to the temperature
dependence of the specific heat below $T_{ca}$. Difference of free
energies of the superfluid and normal phases can be represented as:
\begin{equation}
 F_s-F_n=-\frac{\Delta C_b}{2T_{cb}}\langle
T_c\rangle^2\int\langle|\psi|^4\rangle d^3r,
\end{equation}
where $\Delta C_b$ is the specific heat jump in pure $^3$He. With
account of the terms of the order of $\langle\chi\rangle^2$
$\langle|\psi|^4\rangle=u^2$, then:
\begin{equation}
\frac{1}{T}(C_s-C_n)=\frac{\Delta
C_b}{2T_{cb}}\frac{\partial^2(u^2)}{\partial t^2}.
\end{equation}
For $u(t)$ given by Eqns.~(12),~(13) this expression remains finite
up to $T_{ca}$, but its validity is justified only in a region where
the condition $\langle\chi^2\rangle\ll 1$ is met.

\section{\label{p-wave}P-wave pairing}
For application to the superfluid $^3$He in aerogel the argument of
Section \ref{scalar} has to be reformulated for the p-wave Cooper
pairing. The order parameter in that case is 3$\times$3 complex
matrix $A_{\mu j}$. The first index refers to 3 spin components, and
the second -- to 3 components of the angular momentum of Cooper
pairs. Aerogel interacts with the orbital part of the order
parameter. This interaction is described by the additional term in
the density of free energy:
\begin{equation}
f_{\eta}=-N_{eff}(\eta_{jl}({\bf r})+\kappa_{jl})A_{\mu j}A_{\mu
l}^*.
\end{equation}
Now $\eta_{jl}({\bf r})$ is a real and symmetric 3$\times$3 random
tensor of local anisotropy. The ensemble average
$\langle\eta_{jl}\rangle=0$. The $\textbf{r}$-independent tensor
$\kappa_{jl}$ is introduced for account of possible global
anisotropy. In the present calculations a uniaxial anisotropy will
be assumed only  implicitly for lifting of the orientational
degeneracy of the A-like phase. Within the used approximation the
global anisotropy will not enter results and it will be dropped from
the free energy. $N_{eff}$ is the overall factor, having
dimensionality of density of states, it does not enter the resulting
equations. For the gradient energy we take a simplified isotropic
expression:
\begin{equation}
f_{\nabla}=N_{eff}\xi_s^2\left(\frac{\partial A_{\mu l}}{\partial
x_n} \frac{\partial A^*_{\mu l}}{\partial x_n}\right).
\end{equation}
Both terms (21) and (22) are added to the unperturbed  density of GL
free energy:
\begin{equation}
 f_0=N_{eff}(\tau A_{\mu j}A_{\mu j}^*+
\frac{1}{2}\sum_{s=1}^5 \beta_sI_s)
\end{equation}
where $I_s$ - are the 4-th order invariants in the expansion of the
free energy over $A_{\mu j}$:\cite{VW} $I_1=A_{\mu j}A_{\mu j}A_{\nu
l}^*A_{\nu l}^*$, $I_2=A_{\mu j}A_{\mu j}^*A_{\nu l}A_{\nu l}^*$,
$I_3=A_{\mu j}A_{\nu j}A_{\mu l}^*A_{\nu l}^*$, $I_4=A_{\mu j}A_{\nu
j}^*A_{\nu l}A_{\mu l}^*$, $I_5=A_{\mu j}A_{\nu j}^*A_{\mu l}A_{\nu
l}^*$, and  $\beta_1,...\beta_5$ are phenomenological coefficients.
The resulting GL equations have the form:
\begin{equation}
\frac{\partial f_0}{\partial A^*_{\mu
j}}-\xi_s^2\left(\frac{\partial^2 A_{\mu j}}{\partial
x_n^2}\right) = -A_{\mu l}\eta_{lj}({\bf r}).
\end{equation}

To follow the perturbation procedure of the previous section we
split the order parameter into its average and the fluctuation. In
the B-like phase we search for a solution in the form $A_{\mu
j}=\langle A_{\mu l}\rangle(\delta_{lj}+c_{lj}({\bf r}))$. Here
$\langle A_{\mu j}\rangle=\bar \Delta  R_{\mu j}$,  $R_{\mu j}$ is a
real orthogonal matrix and $c_{lj}({\bf r})$ is real and symmetric.
Procedure analogous to that used at the derivation of
Eq.~(\ref{Ave_Eq}) renders an equation for $\bar\Delta^2$:
\begin{eqnarray}
\bar\Delta^2\{(3\beta_{12}+\beta_{345})+ \beta_{12345}\langle
c_{ml}c_{lm}\rangle\nonumber\\
+\frac{2}{3}\beta_{12}\langle c_{mm}c_{ll}\rangle\}=
\frac{1}{3}\langle\eta_{ml}c_{lm}\rangle-\tau.
\end{eqnarray}
Here $\beta_{12}=\beta_1+\beta_2$ etc.. Tensor of
fluctuations $c_{jl}$ satisfies the linear equation:
\begin{eqnarray}
(\tau\delta_{jn}-\eta_{jn})c_{nl}-\xi_s^2\nabla^2c_{jl}+\bar{\Delta}^2
[3\beta_{12345}c_{jl}\nonumber\\+\beta_{12}2c_{nn}\delta_{jl}]=
\eta_{jl}-\langle\eta_{jn}c_{nl}\rangle. \label{GL-linear}
\end{eqnarray}
As before we consider aerogel as a random distribution of identical
scattering centers. The center situated at a point ${\bf r}_a$
introduces a local
perturbation~\begin{math}\eta_{jl}^{(1)}(\textbf{r}~-~\textbf{r}_a)\end{math}.
For sufficiently dilute system, when effect of multiple scattering
of quasi-particles by impurities can be neglected, the overall
perturbation is a sum of contributions of individual impurities:
$\eta_{jl}({\bf r})=\sum_a\eta_{jl}^{(1)}(\textbf{r}-\textbf{r}_a)$.
Its Fourier transform is:
\begin{eqnarray}
\eta_{jl}({\bf k})=\eta_{jl}^{(1)}({\bf k}) \sum_a\exp{(-i{\bf
k}{\bf r_a})}.
\end{eqnarray}
As has been discussed in Section \ref{scalar} most essential are
fluctuations with small ${\bf k}$, which meet the condition
$k\xi_s\ll 1$. That makes possible to substitute for
$\eta_{jl}^{(1)}({\bf k})$ its limiting value at ${\bf k}\rightarrow
0$ which is $\delta_{jl}\eta^{(1)}(0)$, leaving dependence on ${\bf
k}$ only in the phase factors $\exp{(-i{\bf k}{\bf r_a})}$. In the
coordinate representation it means that the scattering centers are
considered as  isotropic and point-like $\eta_{jl}^{(1)}({\bf
r})=\delta_{jl}\eta^{(1)}(0)\delta(\textbf{r}-\textbf{r}_a)$ The
total tensor $\eta_{jl}({\bf r})$ in that case is also proportional
to $\delta_{jl}$. By the argument of Rainer and Vuorio \cite{Rainer}
$\eta^{(1)}(0)=\gamma(\xi_s/nl)$ where $l$ is a  transport mean free
path and the coefficient $\gamma$ depends on properties of the
centers. For diffusely scattering balls $\gamma=\pi^2/4$.\cite{fom1}
Symmetric tensor $c_{jl}$ in Eq.~(\ref{GL-linear}) can be
represented as a sum of its scalar $c_{nn}\delta_{jl}$ and traceless
$c^{(s)}_{jl}=\frac{1}{2}(c_{jl}+c_{lj}-\frac{2}{3}c_{nn}\delta_{jl})$
parts. The system of equations (\ref{GL-linear}) splits then in two
independent systems for each part of $c_{jl}$. Of the two only
equation for the scalar part contains a finite perturbation in the
right hand side, so that the solution can be sought in a form
$c_{jl}=\delta_{jl}\chi$. Substitution of
$\eta_{jl}=\eta\delta_{jl}$ in Eqns.~(25),~(\ref{GL-linear}) and
introduction of the notation $\langle\psi\rangle^2=\bar\Delta^2
(3\beta_{12}+\beta_{345})$ transforms Eq.~(25) into
Eq.~(\ref{Ave_Eq}) and Eq.~(\ref{GL-linear}) -- into
Eq.~(\ref{Fluc_Eq}). The problem for $^3$He-B turns out to be
equivalent to that for the scalar order parameter.

In the A-like phase situation is complicated by a possibility of
formation of the disordered Larkin-Imry-Ma (LIM) state. Global
anisotropy orients the order parameter and restores the long-range
order. Experiments \cite{knmts} and theoretical estimations
\cite{vol,SF2007} show that compression of aerogel for a few percent
along one direction is sufficient for the restoration of the long
range order. In this case reduction to the scalar problem is
possible for the A-like phase as well. In the bulk liquid the order
parameter of the A-phase is a direct product of spin and orbital
vectors: $A_{\mu j}=d_\mu A_j$. Aerogel coated by $^4$He does not
interact with the spin part of the order parameter. We seek a
solution of Eq.~(24) also in a separable form, then it reduces to
the equation for the orbital vector $A_j$:
\begin{eqnarray}
[\tau\delta_{jl}-\eta_{jl}(\textbf{r})]A_l-\xi_s^2\left(\frac{\partial^2
A_j}{\partial
x_n^2}\right)+\beta_{13}A_j^*(A_sA_s)\nonumber\\
+\beta_{245}A_j(A_sA_s^*)=0
\end{eqnarray}

In line with the discussion in the paragraph following Eq.~(27) we
keep only isotropic part of the perturbation
$\eta_{jl}(\textbf{r})=\delta_{jl}\eta(\textbf{r})$ and seek the
solution in a form $A_l=\langle A_l\rangle+a_l$. Here $\langle
A_l\rangle=(1/\sqrt{2})\bar\Delta(\hat{m}+i\hat{n})$, unit vectors
$\hat{m}$ and $\hat{n}$ are mutually orthogonal and orthogonal to
the direction of compression. For definiteness we have chosen one of
two possible orientations of $\langle A_l\rangle$. Repeating the
procedure of Section \ref{scalar} we find as the solution
$a_l=\chi(\textbf{r})\langle A_l\rangle$. Denoting
$\beta_{245}\bar\Delta^2=\langle\psi\rangle^2$ we return to the
equations (4) and (5) of the scalar problem.

The reduction to the scalar order parameter is based on the fact
that principal contribution to thermodynamic anomalies comes from
inhomogeneities with small wave vectors. Such inhomogeneities
interact with fluctuations of the amplitude of the order parameter,
their coupling with the other collective modes of B-like and A-like
phases appears only in the next order on $u$.

\section{\label{comparison}Comparison with experiment}
\begin{figure}[t]
\includegraphics[scale=0.37]{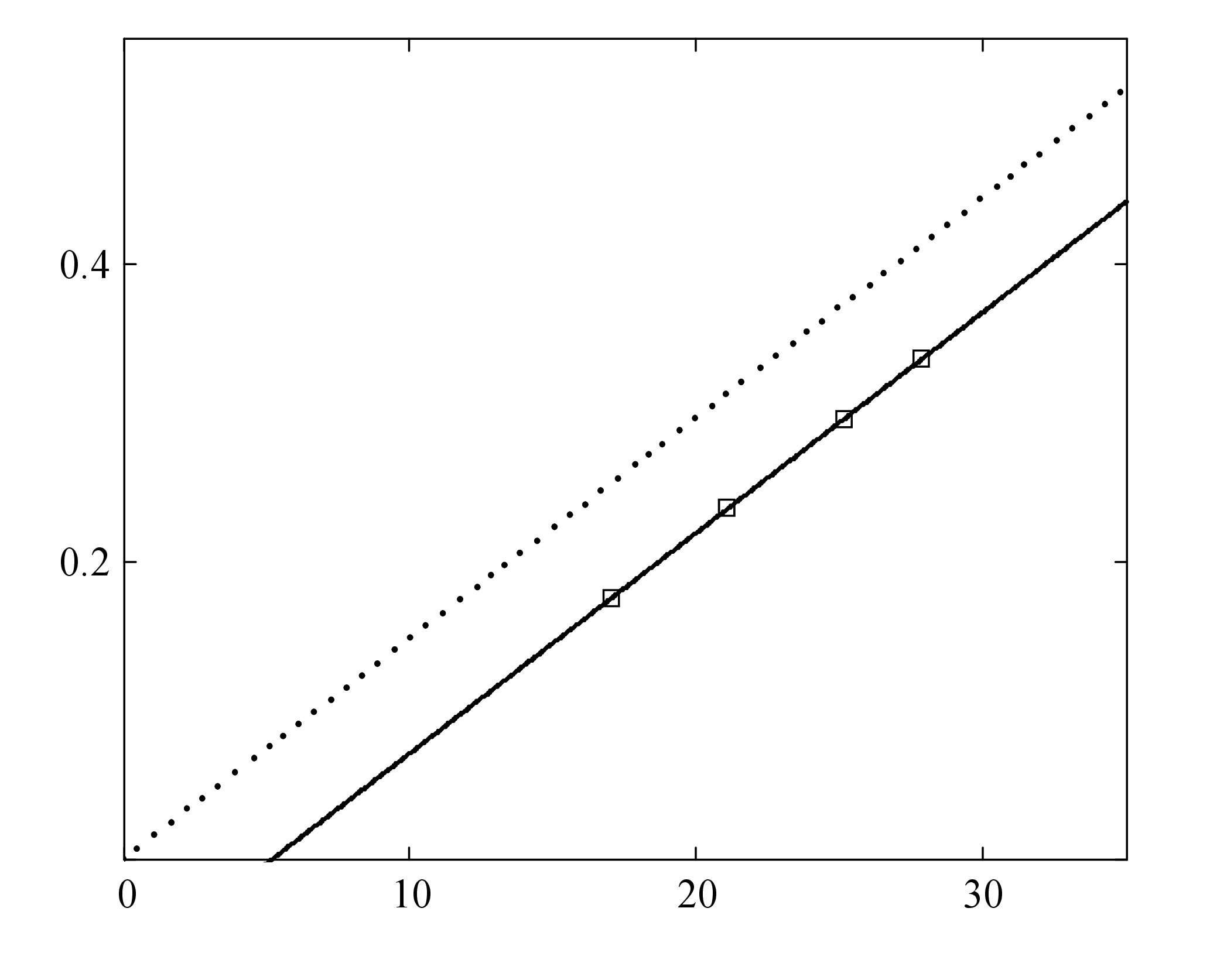}
\scalebox{0.9}[0.9]{\rotatebox{90}{$~~~~~~~~~~~~~~~~~[{(T_{ca}-T_{cb})}/{T_{cb}}](P)$}}
\scalebox{0.9}[0.9]{\rotatebox{0}{$~~~~~~~~~~~~\xi_0(P)~~~~~~~~~~~$}}
\caption{\label{fig:epsart2} Fitting of the dependence of lowering
of the transition temperature of $^3$He in aerogel (data points from
Ref.~\onlinecite{feff}) as a function of $\xi_0=2\pi\hbar
v_F/T_{cb}$ by the straight line $\gamma\xi_0/l+\delta$ (solid
line). The slope $\gamma=2.95$ falls between its values for specular
$\gamma_{sp}=\pi^2/3$ and diffuse scattering $\gamma_{dif}=\pi^2/4$.
The constant shift $\delta=0.075$ is a contribution of correlations
to the change of the transition temperature ($\delta\sim AR^2/l^2$).
The dotted line  corresponds to non-correlated impurities.}
\end{figure}

The obtained expressions allow comparison with experiment.
Properties of aerogel enter these expressions via the mean free path
$l$ and the structure factor  $S(\textbf{k})$. Both quantities in
principle can be found from independent experiments, $l$ -- from the
measurement of spin diffusion in the normal phase,\cite{bun-sauls}
and $S(\textbf{k})$ -- from the X-ray scattering data. Physical
quantities which can be compared with our calculations are e.g. the
temperature dependence of the square of the longitudinal resonance
frequency $\Omega_L^2$, which in a vicinity of $T_c$ is proportional
to $\bar{\Delta}^2$ i.e. $\langle\psi\rangle^2$
(Eqns.~(4),~(12),~(13)) and the lowering of the transition
temperature, given by Eq.~(\ref{trans_temp_shift}). In realization
of the comparison we encounter two difficulties. Firstly the
thermodynamic data suitable for the comparison are not accompanied
by the independently measured $l$ and $S(\textbf{k})$ for the same
sample of aerogel. Even when $S(\textbf{k})$ is measured the result
is given in arbitrary units, leaving ambiguity in the overall
amplitude. Secondly the theoretical expressions are obtained within
the perturbation theory and their application is limited by a region
of small fluctuations. Since the discussed effects originate from
fluctuations they are small in a region of applicability of theory.
So, we have to find a compromise between the magnitude of the effect
and accuracy of its description. With these reservations we choose
for comparison a temperature dependence of $\Omega_L^2$, extracted
from the measurements of the transverse NMR frequency
shift.\cite{dmit4} In these experiments neither $l$ nor
$S(\textbf{k})$ was measured, but the data was taken for several
pressures with the same sample of 98\% aerogel and in a suitable
temperature interval. For such porosity  $l$ is roughly estimated as
130-180 nm.\cite{bun-sauls} Fine tuning of $l$ within this interval
is made in a process of fitting. Unknown structure factor formally
introduces infinitely many fitting parameters, but in all equations
$S(\textbf{k})$ enters under the sign of integral. The main
contribution to the integral in Eq.~(13) comes from the region of
small $\textbf{k}$ so that only the asymptotic form of
$S(\textbf{k})$ at $k\rightarrow 0$ or of $v(r)$ at $r\rightarrow
\infty$ is essential. The integral in Eq.~(\ref{trans_temp_shift})
for $D_f>1$ is also defined by a region of small $\textbf{k}$. The
answers are not very sensitive to the detailed form of
$S(\textbf{k})$ or of $v(r)$, so we use for interpretation of data
the model expression Eq.~(16) with two fitting parameters $A$ and
$R$. Lowering of the transition temperature by aerogel was discussed
before.\cite{fom1} In the leading order on $\xi_0/l$ the principal
contribution is proportional to this ratio and correction due to
correlations depends on the combination $AR^2/l^2$. For comparison
we use the data of Ref.~\onlinecite{dmit4} taken at four pressures:
29.3, 17.5, 20.1 and 11.9 bar. Transition temperatures for all four
pressures can be fitted by a straight line as shown on the
Fig.~\ref{fig:epsart2} and the combination $AR^2/l^2$ can be
extracted from the fit.
\begin{figure}[t]
\includegraphics[scale=0.62]{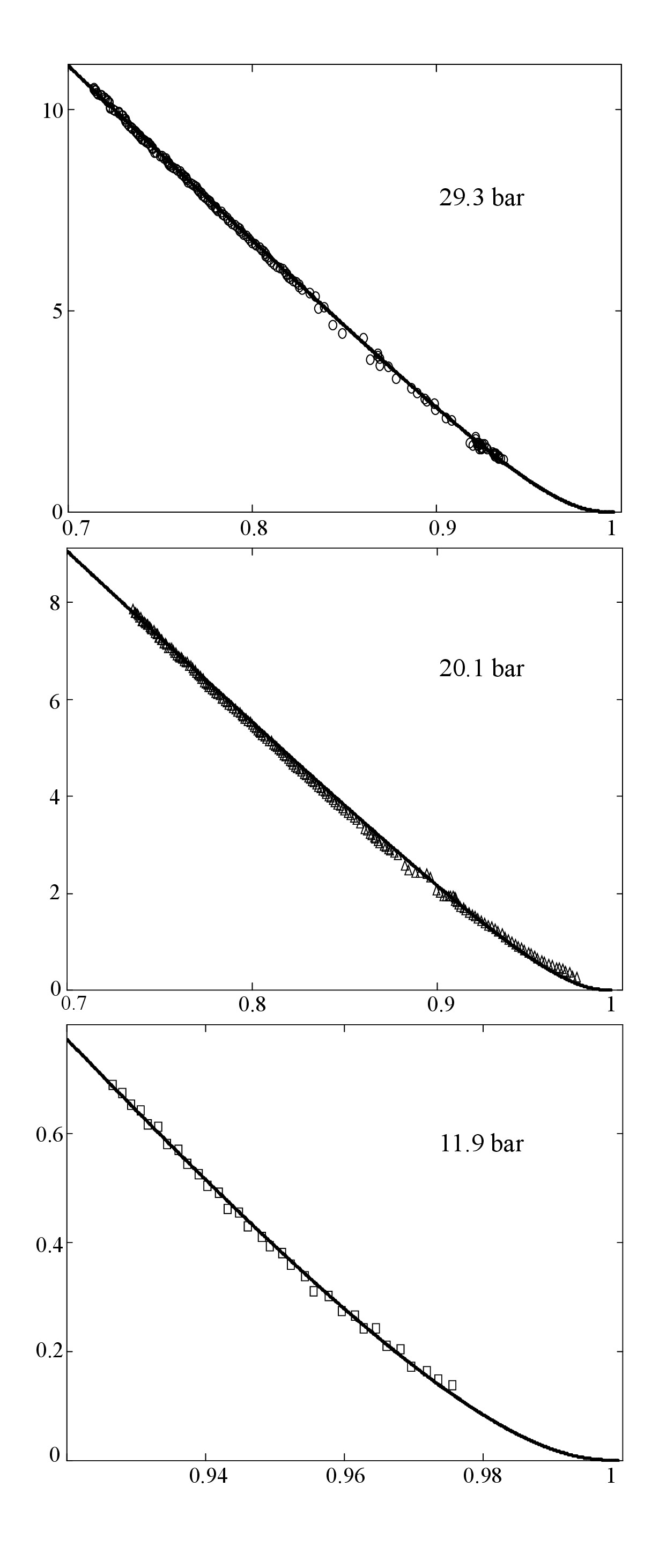}
\scalebox{0.9}[0.9]{\rotatebox{90}{$~~~~~~~~~~~~~~~~~~~~~~~~~~~~~~~~~~~~~~~~~~~~~~~~~~~~~~~~~~~~~~~~~~~~~~~~(\Omega_B/2\pi)^2~[10^9$Hz$^2$]}}
\scalebox{0.9}[0.9]{\rotatebox{0}{$~~~~~~~~~~~~T/T_{ca}~~~~~~~~~~~$}}
\caption{\label{fig:epsart4} Temperature dependence of the square of
the longitudinal resonance frequency $\Omega_B^2$ in the B-like
phase at three different pressures. Data points are taken from
Ref.~\onlinecite{feff}, solid lines -- results of fitting.
Parameters of aerogel are $l=140$~nm, $R=35$~nm, $A=0.055$.
Coherence length is taken as $\xi_s=0.71\xi_0$.}
\end{figure}
Additional combinations of parameters are found by fitting
temperature dependencies of $\Omega_L^2$ in the B-like phase, as
shown on the Fig.~\ref{fig:epsart4}. The procedure requires further
fitting parameters originating from the phenomenological
coefficients, entering Eqns.~(1),~(24). The amplitude $\Psi_0^2$ has
to be fitted for each pressure. About the coherence length $\xi_s$
we assume that it scales as $(\hbar v_F)/T_{cb}$ and fit one
coefficient for all pressures. Parameters of aerogel, which provide
the best fit for the lowering of the transition temperature for all
four pressures and for the temperature dependence of the square of
the longitudinal resonance frequency $\Omega_L^2$ for three
pressures, shown on the Fig.~\ref{fig:epsart2} are the following:
$l=140$ nm, $R=35$ nm, $A=0.055$.

Part of the data fall in a region were $\langle\chi\rangle^2$ is not
very small. Estimation, made with the cited parameters renders for
P=11.4 bar $\langle\chi\rangle^2\approx 0.04/\sqrt{u}$ at
$u\rightarrow 0$. At the position of the closest to $T_c$
experimental point $\langle\chi\rangle^2\approx 1/3$. Fluctuations
are even bigger at P=20.1 bar, where $\langle\chi\rangle^2\approx
0.1/\sqrt{u}$. Argument, based on the perturbation theory can not
apply in a region of developed fluctuations. The argument certainly
breaks down when $\langle\chi\rangle^2\approx 1$. For P=20.1 bar it
would correspond to $u=0.01$.

\begin{figure}[t]
\includegraphics[scale=0.37]{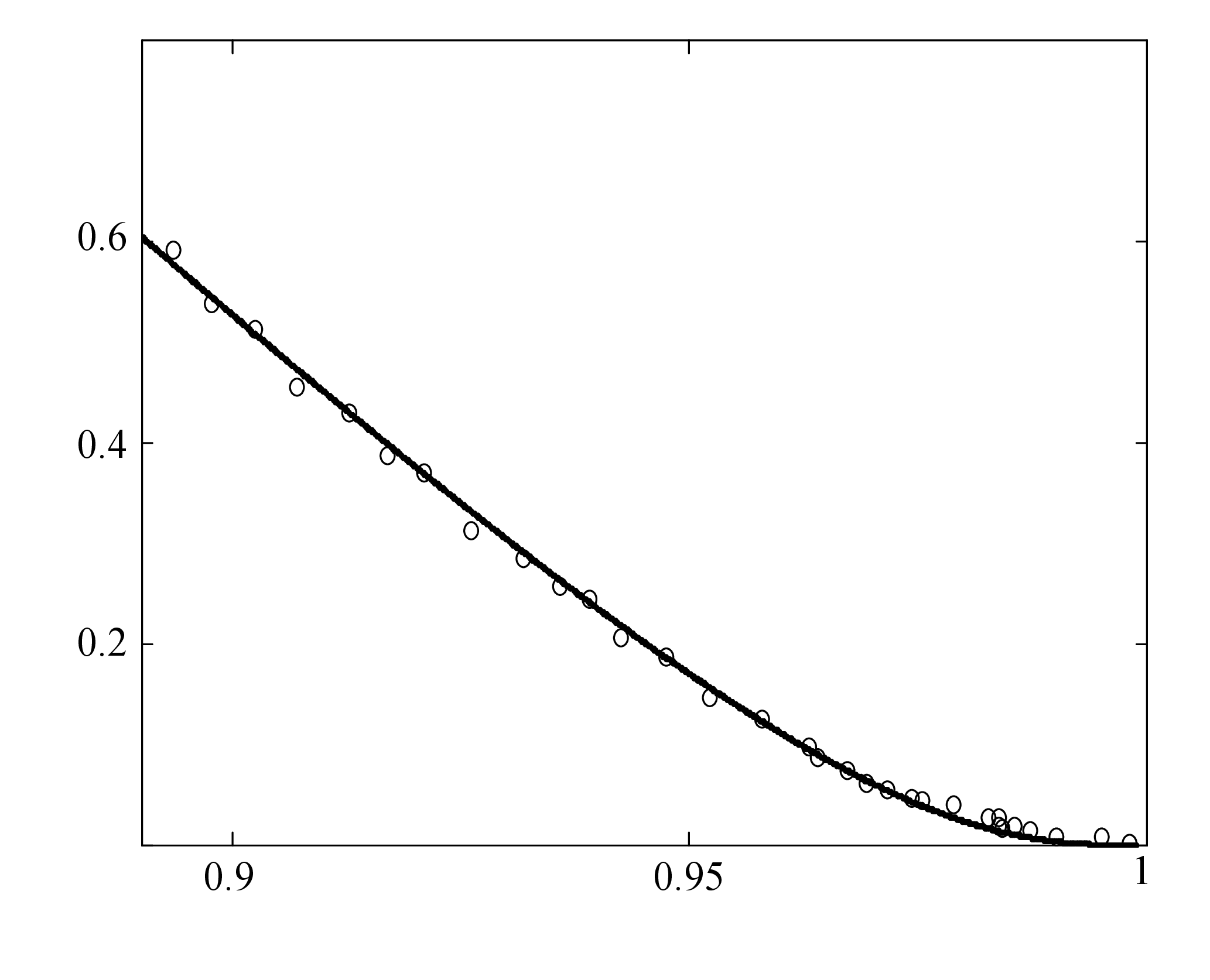}
\scalebox{0.9}[0.9]{\rotatebox{90}{$~~~~~~~~~~~~~~~~~(\Omega_A/2\pi)^2~[10^8$Hz$^2$]}}
\scalebox{0.9}[0.9]{\rotatebox{0}{$~~~~~~~~~~~~T/T_{ca}~~~~~~~~~~~$}}
\caption{\label{fig:epsart3} Temperature dependence of the square of
the longitudinal resonance frequency $\Omega_A^2$ in the A-like
phase at P=29.3 bar. Data points are taken from
Ref.~\onlinecite{dmit4}, solid line -- results of fitting with
parameters of aerogel -- $l=140$ nm, $R=35$ nm, $A=0.055$. Coherence
length is taken as $\xi_s=0.71\xi_0$.}
\end{figure}

For the A-like phase we also use the NMR data of
Ref.~\onlinecite{dmit4}, taken in a condition when compression is
sufficiently strong for orientation of the average order parameter.
Fig.~\ref{fig:epsart3} demonstrates that a satisfactory agreement
with experiment can be achieved.

The results of comparison show that the data for both A-like and B-like phases can be
fitted with the values of parameters which are within the range admitted by other
experiments with different samples of 98\% aerogel. That provides a support to the
suggested interpretation of the observed anomalies of thermodynamic properties of the
superfluid  $^3$He in aerogel.

\section{Conclusion}
When discussing reaction of a superconductor or a superfluid on a
quenched disorder it is important to take into consideration that
reaction of the condensate of Cooper pairs increases with a decrease
of a wave vector of a perturbation  $\textbf{k}$, saturating at
$k\sim 1/\xi(u)$. Correlations between impurities influence
distribution of fluctuations over their Fourier spectra. If, like in
aerogel, correlations increase the weight of the long wavelength
fluctuations effect of the disorder can be significantly increased
in comparison with the uniform spectral distribution of the same
impurities. This results in the increase of a temperature interval
below $T_c$ where fluctuations dominate. The fluctuation region has
a ``tail'', in which fluctuations of the order parameter are
sufficiently small to be treated by the perturbation theory, but
their effect on thermodynamic properties of the superfluid is still
distinguishable. A length of the tail  depends on the relation
between the coherence length of the condensate $\xi_0$ and
correlation radius $R$. For $R\approx\xi_0$ a trace of fluctuations
extends for the most of the Ginzburg and Landau region.

Shift of fluctuations in a region of small $k$ enhances the
transition temperature of $^3$He (cf. Eq.~(\ref{trans_temp_shift})).
Analogous effect was discussed before as a way to further enhance
the $T_c$ in high temperature superconductors  by ordering of
impurities or modulation of paring interaction.\cite{liu,kivel}
Advantage of $^3$He in aerogel as an object of investigation is that
aerogel forms a rigid structure, which can be studied separately
from $^3$He.

\begin{acknowledgments}
We thank V. V. Dmitriev for providing us with the detailed data for
the NMR frequency shift and stimulating discussions and J. M. Parpia
for instructive and encouraging discussions. This research was
supported in part by the Russian Foundation for Basic Research under
projects \#11-02-00357-a, \#11-02-12069-ofi-m-2011
\end{acknowledgments}

\bibliography{liter1}

\providecommand{\noopsort}[1]{}\providecommand{\singleletter}[1]{#1}%
\begin{thebibliography}{24}%
\makeatletter
\providecommand \@ifxundefined [1]{%
 \@ifx{#1\undefined}
}%
\providecommand \@ifnum [1]{%
 \ifnum #1\expandafter \@firstoftwo
 \else \expandafter \@secondoftwo
 \fi
}%
\providecommand \@ifx [1]{%
 \ifx #1\expandafter \@firstoftwo
 \else \expandafter \@secondoftwo
 \fi
}%
\providecommand \natexlab [1]{#1}%
\providecommand \enquote  [1]{``#1''}%
\providecommand \bibnamefont  [1]{#1}%
\providecommand \bibfnamefont [1]{#1}%
\providecommand \citenamefont [1]{#1}%
\providecommand \href@noop [0]{\@secondoftwo}%
\providecommand \href [0]{\begingroup \@sanitize@url \@href}%
\providecommand \@href[1]{\@@startlink{#1}\@@href}%
\providecommand \@@href[1]{\endgroup#1\@@endlink}%
\providecommand \@sanitize@url [0]{\catcode `\\12\catcode `\$12\catcode
  `\&12\catcode `\#12\catcode `\^12\catcode `\_12\catcode `\%12\relax}%
\providecommand \@@startlink[1]{}%
\providecommand \@@endlink[0]{}%
\providecommand \url  [0]{\begingroup\@sanitize@url \@url }%
\providecommand \@url [1]{\endgroup\@href {#1}{\urlprefix }}%
\providecommand \urlprefix  [0]{URL }%
\providecommand \Eprint [0]{\href }%
\providecommand \doibase [0]{http://dx.doi.org/}%
\providecommand \selectlanguage [0]{\@gobble}%
\providecommand \bibinfo  [0]{\@secondoftwo}%
\providecommand \bibfield  [0]{\@secondoftwo}%
\providecommand \translation [1]{[#1]}%
\providecommand \BibitemOpen [0]{}%
\providecommand \bibitemStop [0]{}%
\providecommand \bibitemNoStop [0]{.\EOS\space}%
\providecommand \EOS [0]{\spacefactor3000\relax}%
\providecommand \BibitemShut  [1]{\csname bibitem#1\endcsname}%
\let\auto@bib@innerbib\@empty
\bibitem [{\citenamefont {Abrikosov}\ and\ \citenamefont {Gorkov}(1961)}]{AG}%
  \BibitemOpen
  \bibfield  {author} {\bibinfo {author} {\bibfnamefont {A.~A.}\ \bibnamefont
  {Abrikosov}}\ and\ \bibinfo {author} {\bibfnamefont {L.~P.}\ \bibnamefont
  {Gorkov}},\ }\href@noop {} {\bibfield  {journal} {\bibinfo  {journal} {Zh.
  Eksp. Teor. Fiz.}\ }\textbf {\bibinfo {volume} {39}},\ \bibinfo {pages}
  {1781} (\bibinfo {year} {1961})},\ \translation{Sov. Phys. JETP \textbf{12},
  1243 (1961)}\BibitemShut {NoStop}%
\bibitem [{\citenamefont {Porto}\ and\ \citenamefont {Parpia}(1995)}]{parp1}%
  \BibitemOpen
  \bibfield  {author} {\bibinfo {author} {\bibfnamefont {J.~V.}\ \bibnamefont
  {Porto}}\ and\ \bibinfo {author} {\bibfnamefont {J.~M.}\ \bibnamefont
  {Parpia}},\ }\href@noop {} {\bibfield  {journal} {\bibinfo  {journal} {Phys.
  Rev. Lett.}\ }\textbf {\bibinfo {volume} {74}},\ \bibinfo {pages} {4667}
  (\bibinfo {year} {1995})}\BibitemShut {NoStop}%
\bibitem [{\citenamefont {Sprague}\ \emph {et~al.}(1995)\citenamefont
  {Sprague}, \citenamefont {Haard}, \citenamefont {Kycia}, \citenamefont
  {Rand}, \citenamefont {Lee}, \citenamefont {Hamot},\ and\ \citenamefont
  {Halperin}}]{halp1}%
  \BibitemOpen
  \bibfield  {author} {\bibinfo {author} {\bibfnamefont {D.~T.}\ \bibnamefont
  {Sprague}}, \bibinfo {author} {\bibfnamefont {T.~M.}\ \bibnamefont {Haard}},
  \bibinfo {author} {\bibfnamefont {J.~B.}\ \bibnamefont {Kycia}}, \bibinfo
  {author} {\bibfnamefont {V.~R.}\ \bibnamefont {Rand}}, \bibinfo {author}
  {\bibfnamefont {Y.}~\bibnamefont {Lee}}, \bibinfo {author} {\bibfnamefont
  {P.}~\bibnamefont {Hamot}}, \ and\ \bibinfo {author} {\bibfnamefont {W.~P.}\
  \bibnamefont {Halperin}},\ }\href@noop {} {\bibfield  {journal} {\bibinfo
  {journal} {Phys. Rev. Lett. C}\ }\textbf {\bibinfo {volume} {75}},\ \bibinfo
  {pages} {661} (\bibinfo {year} {1995})}\BibitemShut {NoStop}%
\bibitem [{\citenamefont {Thuneberg}\ \emph {et~al.}(1998)\citenamefont
  {Thuneberg}, \citenamefont {Yip}, \citenamefont {Fogelstrom},\ and\
  \citenamefont {Sauls}}]{thuneb}%
  \BibitemOpen
  \bibfield  {author} {\bibinfo {author} {\bibfnamefont {E.~V.}\ \bibnamefont
  {Thuneberg}}, \bibinfo {author} {\bibfnamefont {S.-K.}\ \bibnamefont {Yip}},
  \bibinfo {author} {\bibfnamefont {M.}~\bibnamefont {Fogelstrom}}, \ and\
  \bibinfo {author} {\bibfnamefont {J.~A.}\ \bibnamefont {Sauls}},\ }\href@noop
  {} {\bibfield  {journal} {\bibinfo  {journal} {Phys. Rev. Lett.}\ }\textbf
  {\bibinfo {volume} {80}},\ \bibinfo {pages} {2861} (\bibinfo {year}
  {1998})}\BibitemShut {NoStop}%
\bibitem [{\citenamefont {Halperin}\ \emph {et~al.}(2008)\citenamefont
  {Halperin}, \citenamefont {Choi}, \citenamefont {Davis},\ and\ \citenamefont
  {Polanen}}]{halp2}%
  \BibitemOpen
  \bibfield  {author} {\bibinfo {author} {\bibfnamefont {W.~P.}\ \bibnamefont
  {Halperin}}, \bibinfo {author} {\bibfnamefont {H.}~\bibnamefont {Choi}},
  \bibinfo {author} {\bibfnamefont {J.~P.}\ \bibnamefont {Davis}}, \ and\
  \bibinfo {author} {\bibfnamefont {J.}~\bibnamefont {Polanen}},\ }\href@noop
  {} {\bibfield  {journal} {\bibinfo  {journal} {J. Phys. Soc. Jpn.}\ }\textbf
  {\bibinfo {volume} {77}},\ \bibinfo {pages} {111002} (\bibinfo {year}
  {2008})}\BibitemShut {NoStop}%
\bibitem [{\citenamefont {Parpia}\ \emph {et~al.}(2008)\citenamefont {Parpia},
  \citenamefont {Fefferman}, \citenamefont {Porto}, \citenamefont {Dmitriev},
  \citenamefont {Levitin},\ and\ \citenamefont {Zmeev}}]{feff}%
  \BibitemOpen
  \bibfield  {author} {\bibinfo {author} {\bibfnamefont {J.~M.}\ \bibnamefont
  {Parpia}}, \bibinfo {author} {\bibfnamefont {A.~D.}\ \bibnamefont
  {Fefferman}}, \bibinfo {author} {\bibfnamefont {J.~V.}\ \bibnamefont
  {Porto}}, \bibinfo {author} {\bibfnamefont {V.~V.}\ \bibnamefont {Dmitriev}},
  \bibinfo {author} {\bibfnamefont {L.~V.}\ \bibnamefont {Levitin}}, \ and\
  \bibinfo {author} {\bibfnamefont {D.~E.}\ \bibnamefont {Zmeev}},\ }\href@noop
  {} {\bibfield  {journal} {\bibinfo  {journal} {J. Low Temp.Phys.}\ }\textbf
  {\bibinfo {volume} {150}},\ \bibinfo {pages} {464} (\bibinfo {year}
  {2008})}\BibitemShut {NoStop}%
\bibitem [{\citenamefont {Porto}\ and\ \citenamefont {Parpia}(1999)}]{parp2}%
  \BibitemOpen
  \bibfield  {author} {\bibinfo {author} {\bibfnamefont {J.~V.}\ \bibnamefont
  {Porto}}\ and\ \bibinfo {author} {\bibfnamefont {J.~M.}\ \bibnamefont
  {Parpia}},\ }\href@noop {} {\bibfield  {journal} {\bibinfo  {journal} {Phys.
  Rev. B}\ }\textbf {\bibinfo {volume} {59}},\ \bibinfo {pages} {14583}
  (\bibinfo {year} {1999})}\BibitemShut {NoStop}%
\bibitem [{\citenamefont {Haard}\ \emph {et~al.}(2000)\citenamefont {Haard},
  \citenamefont {Gervais}, \citenamefont {Nomura},\ and\ \citenamefont
  {Halperin}}]{haard}%
  \BibitemOpen
  \bibfield  {author} {\bibinfo {author} {\bibfnamefont {T.~M.}\ \bibnamefont
  {Haard}}, \bibinfo {author} {\bibfnamefont {G.}~\bibnamefont {Gervais}},
  \bibinfo {author} {\bibfnamefont {R.}~\bibnamefont {Nomura}}, \ and\ \bibinfo
  {author} {\bibfnamefont {W.~P.}\ \bibnamefont {Halperin}},\ }\href@noop {}
  {\bibfield  {journal} {\bibinfo  {journal} {Physica B}\ }\textbf {\bibinfo
  {volume} {284-288}},\ \bibinfo {pages} {289} (\bibinfo {year}
  {2000})}\BibitemShut {NoStop}%
\bibitem [{\citenamefont {Hanninen}\ and\ \citenamefont
  {Thuneberg}(2003)}]{han}%
  \BibitemOpen
  \bibfield  {author} {\bibinfo {author} {\bibfnamefont {R.}~\bibnamefont
  {Hanninen}}\ and\ \bibinfo {author} {\bibfnamefont {E.~V.}\ \bibnamefont
  {Thuneberg}},\ }\href@noop {} {\bibfield  {journal} {\bibinfo  {journal}
  {Phys. Rev. B}\ }\textbf {\bibinfo {volume} {67}},\ \bibinfo {pages} {214507}
  (\bibinfo {year} {2003})}\BibitemShut {NoStop}%
\bibitem [{\citenamefont {Sauls}\ and\ \citenamefont {Sharma}(2003)}]{sauls2}%
  \BibitemOpen
  \bibfield  {author} {\bibinfo {author} {\bibfnamefont {J.~A.}\ \bibnamefont
  {Sauls}}\ and\ \bibinfo {author} {\bibfnamefont {P.}~\bibnamefont {Sharma}},\
  }\href@noop {} {\bibfield  {journal} {\bibinfo  {journal} {Phys. Rev. B}\
  }\textbf {\bibinfo {volume} {68}},\ \bibinfo {pages} {224502} (\bibinfo
  {year} {2003})}\BibitemShut {NoStop}%
\bibitem [{\citenamefont {Larkin}\ and\ \citenamefont
  {Ovchinnikov}(1971)}]{LO}%
  \BibitemOpen
  \bibfield  {author} {\bibinfo {author} {\bibfnamefont {A.~I.}\ \bibnamefont
  {Larkin}}\ and\ \bibinfo {author} {\bibfnamefont {Y.~N.}\ \bibnamefont
  {Ovchinnikov}},\ }\href@noop {} {\bibfield  {journal} {\bibinfo  {journal}
  {Zh. Eksp. Teor. Fiz.}\ }\textbf {\bibinfo {volume} {61}},\ \bibinfo {pages}
  {1221} (\bibinfo {year} {1971})},\ \translation{Sov. Phys. JETP \textbf{34},
  651 (1971)}\BibitemShut {NoStop}%
\bibitem [{\citenamefont {Feldman}(2001)}]{feldm}%
  \BibitemOpen
  \bibfield  {author} {\bibinfo {author} {\bibfnamefont {D.~E.}\ \bibnamefont
  {Feldman}},\ }\href@noop {} {\bibfield  {journal} {\bibinfo  {journal} {Int.
  Journ. Mod. Phys.}\ }\textbf {\bibinfo {volume} {15}},\ \bibinfo {pages}
  {2954} (\bibinfo {year} {2001})}\BibitemShut {NoStop}%
\bibitem [{\citenamefont {Fomin}(2008)}]{fom1}%
  \BibitemOpen
  \bibfield  {author} {\bibinfo {author} {\bibfnamefont {I.~A.}\ \bibnamefont
  {Fomin}},\ }\href@noop {} {\bibfield  {journal} {\bibinfo  {journal} {Pis'ma
  v ZhETF}\ }\textbf {\bibinfo {volume} {88}},\ \bibinfo {pages} {65} (\bibinfo
  {year} {2008})},\ \translation{JETP Letters \textbf{88}, 59
  (2008)}\BibitemShut {NoStop}%
\bibitem [{\citenamefont {Fomin}(2011)}]{fom2}%
  \BibitemOpen
  \bibfield  {author} {\bibinfo {author} {\bibfnamefont {I.~A.}\ \bibnamefont
  {Fomin}},\ }\href@noop {} {\bibfield  {journal} {\bibinfo  {journal} {Pis'ma
  v ZhETF}\ }\textbf {\bibinfo {volume} {93}},\ \bibinfo {pages} {159}
  (\bibinfo {year} {2011})},\ \translation{JETP Letters \textbf{93}, 144
  (2011)}\BibitemShut {NoStop}%
\bibitem [{\citenamefont {Freltoft}\ \emph {et~al.}(1986)\citenamefont
  {Freltoft}, \citenamefont {Kjems},\ and\ \citenamefont {Sinha}}]{frel}%
  \BibitemOpen
  \bibfield  {author} {\bibinfo {author} {\bibfnamefont {T.}~\bibnamefont
  {Freltoft}}, \bibinfo {author} {\bibfnamefont {J.~K.}\ \bibnamefont {Kjems}},
  \ and\ \bibinfo {author} {\bibfnamefont {S.~K.}\ \bibnamefont {Sinha}},\
  }\href@noop {} {\bibfield  {journal} {\bibinfo  {journal} {Phys. Rev. B}\
  }\textbf {\bibinfo {volume} {33}},\ \bibinfo {pages} {269} (\bibinfo {year}
  {1986})}\BibitemShut {NoStop}%
\bibitem [{\citenamefont {Vollhardt}\ and\ \citenamefont {Woelfle}(1990)}]{VW}%
  \BibitemOpen
  \bibfield  {author} {\bibinfo {author} {\bibfnamefont {D.}~\bibnamefont
  {Vollhardt}}\ and\ \bibinfo {author} {\bibfnamefont {P.}~\bibnamefont
  {Woelfle}},\ }\href@noop {} {\emph {\bibinfo {title} {The Superfluid Phases
  of Helium 3}}}\ (\bibinfo  {publisher} {Tailor and Francis},\ \bibinfo {year}
  {1990})\BibitemShut {NoStop}%
\bibitem [{\citenamefont {Rainer}\ and\ \citenamefont {Vuorio}(1977)}]{Rainer}%
  \BibitemOpen
  \bibfield  {author} {\bibinfo {author} {\bibfnamefont {D.}~\bibnamefont
  {Rainer}}\ and\ \bibinfo {author} {\bibfnamefont {M.}~\bibnamefont
  {Vuorio}},\ }\href@noop {} {\bibfield  {journal} {\bibinfo  {journal} {J.
  Phys. C: Solid State Phys.}\ }\textbf {\bibinfo {volume} {10}},\ \bibinfo
  {pages} {3093} (\bibinfo {year} {1977})}\BibitemShut {NoStop}%
\bibitem [{\citenamefont {Kunimatsu}\ \emph {et~al.}(2007)\citenamefont
  {Kunimatsu}, \citenamefont {Sato}, \citenamefont {Izumina}, \citenamefont
  {Matsubara}, \citenamefont {Sasaki}, \citenamefont {Kubota}, \citenamefont
  {ishikawa}, \citenamefont {Mizusaki},\ and\ \citenamefont {Bunkov}}]{knmts}%
  \BibitemOpen
  \bibfield  {author} {\bibinfo {author} {\bibfnamefont {T.}~\bibnamefont
  {Kunimatsu}}, \bibinfo {author} {\bibfnamefont {T.}~\bibnamefont {Sato}},
  \bibinfo {author} {\bibfnamefont {K.}~\bibnamefont {Izumina}}, \bibinfo
  {author} {\bibfnamefont {A.}~\bibnamefont {Matsubara}}, \bibinfo {author}
  {\bibfnamefont {Y.}~\bibnamefont {Sasaki}}, \bibinfo {author} {\bibnamefont
  {Kubota}}, \bibinfo {author} {\bibfnamefont {O.}~\bibnamefont {ishikawa}},
  \bibinfo {author} {\bibfnamefont {T.}~\bibnamefont {Mizusaki}}, \ and\
  \bibinfo {author} {\bibfnamefont {Y.~M.}\ \bibnamefont {Bunkov}},\
  }\href@noop {} {\bibfield  {journal} {\bibinfo  {journal} {Pis'ma v ZhETF}\
  }\textbf {\bibinfo {volume} {86}},\ \bibinfo {pages} {244} (\bibinfo {year}
  {2007})},\ \translation{JETP Letters \textbf{86}, 216 (2007)}\BibitemShut
  {NoStop}%
\bibitem [{\citenamefont {Volovik}(2008)}]{vol}%
  \BibitemOpen
  \bibfield  {author} {\bibinfo {author} {\bibfnamefont {G.~E.}\ \bibnamefont
  {Volovik}},\ }\href@noop {} {\bibfield  {journal} {\bibinfo  {journal} {J.
  Low Temp. Phys.}\ }\textbf {\bibinfo {volume} {150}},\ \bibinfo {pages} {453}
  (\bibinfo {year} {2008})}\BibitemShut {NoStop}%
\bibitem [{\citenamefont {Surovtsev}\ and\ \citenamefont
  {Fomin}(2008)}]{SF2007}%
  \BibitemOpen
  \bibfield  {author} {\bibinfo {author} {\bibfnamefont {E.~V.}\ \bibnamefont
  {Surovtsev}}\ and\ \bibinfo {author} {\bibfnamefont {I.~A.}\ \bibnamefont
  {Fomin}},\ }\href@noop {} {\bibfield  {journal} {\bibinfo  {journal} {J. Low
  Temp. Phys.}\ }\textbf {\bibinfo {volume} {150}},\ \bibinfo {pages} {487}
  (\bibinfo {year} {2008})}\BibitemShut {NoStop}%
\bibitem [{\citenamefont {Sauls}\ \emph {et~al.}(2005)\citenamefont {Sauls},
  \citenamefont {Bunkov}, \citenamefont {Collin}, \citenamefont {Godfrin},\
  and\ \citenamefont {Sharma}}]{bun-sauls}%
  \BibitemOpen
  \bibfield  {author} {\bibinfo {author} {\bibfnamefont {J.~A.}\ \bibnamefont
  {Sauls}}, \bibinfo {author} {\bibfnamefont {Y.~M.}\ \bibnamefont {Bunkov}},
  \bibinfo {author} {\bibfnamefont {E.}~\bibnamefont {Collin}}, \bibinfo
  {author} {\bibfnamefont {H.}~\bibnamefont {Godfrin}}, \ and\ \bibinfo
  {author} {\bibfnamefont {P.}~\bibnamefont {Sharma}},\ }\href@noop {}
  {\bibfield  {journal} {\bibinfo  {journal} {Phys. Rev. B}\ }\textbf {\bibinfo
  {volume} {72}},\ \bibinfo {pages} {024507} (\bibinfo {year}
  {2005})}\BibitemShut {NoStop}%
\bibitem [{\citenamefont {Dmitriev}\ \emph {et~al.}(2007)\citenamefont
  {Dmitriev}, \citenamefont {Krasnikhin}, \citenamefont {Mulders},
  \citenamefont {Zavjalov},\ and\ \citenamefont {Zmeev}}]{dmit4}%
  \BibitemOpen
  \bibfield  {author} {\bibinfo {author} {\bibfnamefont {V.~V.}\ \bibnamefont
  {Dmitriev}}, \bibinfo {author} {\bibfnamefont {D.~A.}\ \bibnamefont
  {Krasnikhin}}, \bibinfo {author} {\bibfnamefont {N.}~\bibnamefont {Mulders}},
  \bibinfo {author} {\bibfnamefont {V.~V.}\ \bibnamefont {Zavjalov}}, \ and\
  \bibinfo {author} {\bibfnamefont {D.~E.}\ \bibnamefont {Zmeev}},\ }\href@noop
  {} {\bibfield  {journal} {\bibinfo  {journal} {Pis'ma v ZhETF}\ }\textbf
  {\bibinfo {volume} {86}},\ \bibinfo {pages} {681} (\bibinfo {year} {2007})},\
  \translation{JETP Letters \textbf{86}, 594 (2007)}\BibitemShut {NoStop}%
\bibitem [{\citenamefont {Liu}\ \emph {et~al.}(2006)\citenamefont {Liu},
  \citenamefont {Yang}, \citenamefont {Qin}, \citenamefont {Yu}, \citenamefont
  {Yang}, \citenamefont {Li}, \citenamefont {Yu}, \citenamefont {Jin},\ and\
  \citenamefont {Uchida}}]{liu}%
  \BibitemOpen
  \bibfield  {author} {\bibinfo {author} {\bibfnamefont {Q.~Q.}\ \bibnamefont
  {Liu}}, \bibinfo {author} {\bibfnamefont {H.}~\bibnamefont {Yang}}, \bibinfo
  {author} {\bibfnamefont {X.~M.}\ \bibnamefont {Qin}}, \bibinfo {author}
  {\bibfnamefont {Y.}~\bibnamefont {Yu}}, \bibinfo {author} {\bibfnamefont
  {L.~X.}\ \bibnamefont {Yang}}, \bibinfo {author} {\bibfnamefont {F.~Y.}\
  \bibnamefont {Li}}, \bibinfo {author} {\bibfnamefont {R.~C.}\ \bibnamefont
  {Yu}}, \bibinfo {author} {\bibfnamefont {C.~Q.}\ \bibnamefont {Jin}}, \ and\
  \bibinfo {author} {\bibfnamefont {S.}~\bibnamefont {Uchida}},\ }\href@noop {}
  {\bibfield  {journal} {\bibinfo  {journal} {Phys. Rev. B}\ }\textbf {\bibinfo
  {volume} {74}},\ \bibinfo {pages} {100506(R)} (\bibinfo {year}
  {2006})}\BibitemShut {NoStop}%
\bibitem [{\citenamefont {Martin}\ \emph {et~al.}(2005)\citenamefont {Martin},
  \citenamefont {Podolsky},\ and\ \citenamefont {Kivelson}}]{kivel}%
  \BibitemOpen
  \bibfield  {author} {\bibinfo {author} {\bibfnamefont {I.}~\bibnamefont
  {Martin}}, \bibinfo {author} {\bibfnamefont {D.}~\bibnamefont {Podolsky}}, \
  and\ \bibinfo {author} {\bibfnamefont {S.~A.}\ \bibnamefont {Kivelson}},\
  }\href@noop {} {\bibfield  {journal} {\bibinfo  {journal} {Phys. Rev. B}\
  }\textbf {\bibinfo {volume} {72}},\ \bibinfo {pages} {060502(R)} (\bibinfo
  {year} {2005})}\BibitemShut {NoStop}%
\end{thebibliography}%

\end{document}